\documentclass{mem}
\usepackage{natbib}
\usepackage{txfonts}
\usepackage{balance}
\usepackage{graphicx}
\usepackage{flushend}
\usepackage[a4paper,breaklinks,pdftex]{hyperref}
\usepackage{xcolor}
\hypersetup{
    colorlinks,
    linkcolor={red!50!black},
    citecolor={blue!50!black},
    urlcolor={blue!80!black}
}

\idline{75}{282}
\begin{document}
\def\teff{$T\rm_{eff }$}
\def\kms{$\mathrm {km s}^{-1}$}

\title{Stochastic Recurrent Neural Networks for Modelling Astronomical Time Series: Advantages and Limitations}

   \subtitle{}

\author{
Xinyue Sheng\inst{1} 
\and Matt Nicholl\inst{1}
\and Nicholas Ross\inst{2}
          }

\institute{
Institute of Gravitational Wave Astronomy and School of Physics and Astronomy, University of Birmingham, Birmingham B15 2TT, UK
\and
Advanced Research Division, Niparo, 41 Dundas Street, Edinburgh,   EH3 6QQ, UK \\
\email{xysheng@star.sr.bham.ac.uk}
}

\authorrunning{Xinyue Sheng}

\titlerunning{SRNN \& AGN for LSST}

\date{Received: Day Month Year; Accepted: Day Month Year}

\abstract{
This paper reviews the Stochastic Recurrent Neural Network (SRNN) as applied to the light curves of Active Galactic Nuclei by \citet{Sheng2022}. 
Astronomical data have inherent limitations arising from telescope capabilities, cadence strategies, inevitable observing weather conditions, and current understanding of celestial objects. 
When applying machine learning methods, it is vital to understand the effects of data limitations on our analysis and ability to make inferences. We take \citet{Sheng2022} as a case study, and illustrate the problems and limitations encountered in implementing the SRNN for simulating AGN variability as seen by the Rubin Observatory.

}
\maketitle{}

\section{Introduction}

Machine learning has become increasingly popular in many branches of astronomical research. In particular, these methods are being applied to the large data sets from wide-field sky surveys. Despite many successes, it is vital to understand that any machine learning implementation comes with difficulties and limitations, and may not always be appropriate for every problem \citep{Kremer2017}.

Various model architectures, such as the many flavours of neural networks, extract features through multiple layers. This is particularly suited to imaging data, though without care these models may extract noise as well as or instead of real astronomical features. However, they are often treated as intelligent ``black boxes" whose complex nonlinear computations prevent researchers from directly understanding the feature extraction process, especially for non-image data. Additionally, methods used for data pre-processing and argumentation can greatly effect the training process and model accuracy \citep{MAHARANA202291}. 

Limitations also inevitably arise from the data themselves. 
For example, there is often a trade off between purity and completeness \citep{Smethurst2021}; simulated data may not be representative of reality; and the quality of the data could be influenced by observing conditions, signal-to-noise ratio (SNR), etc. 
It is becoming more crucial to understand these effects with increasingly big data from large sky surveys, such as Vera Rubin Observatory Legacy Survey of Space and Time (LSST), where the utility of the data for a particular investigation are also influenced by the chosen filter and cadence strategies. 

In this paper, we summarize and review the paper \citet{Sheng2022}, discussing the advantages and limitations of applying machine learning techniques for astronomical research purposes using a novel neural network architecture -- a stochastic recurrent neural network (SRNN) -- as a case study. To our knowledge, this was the first application of the SRNN in astronomical research.
In that project, the motivation was to estimate the suitability of different cadence strategies in the upcoming LSST survey for studying the variability in active galactic nuclei (AGN) time series.
The SRNN was applied model simulated AGN light curves as observed with various proposed cadence strategies for the LSST Wide-Fast-Deep (WFD) survey \citep{LSST_Collab2009}.

\section{Data sets}
To evaluate LSST cadences over a 10-year observation period, AGN light curves are simulated using Continuous Auto-Regressive Moving Average (CARMA) models. 

The CARMA model is a statistical description of stochastic and stationary processes in time series. Although it is not a physical model, it has been widely employed as a description of long-term AGN variability.
CARMA models are notated as CARMA(p, q) where p gives the order of the Autoregressive (AR) process and q gives the description of the Moving Average (MA) process.
The first order CARMA model -- CARMA(1,0) or the Damped Random Walk (DRW) -- has been applied to many quasar variability studies \citep[e.g.][]{Kelly2014, Feigelson2018, Moreno2019}. 
It can be expressed by Equation \ref{eq: drw}, 
where $\alpha$ is the C-AR coefficient and $\beta$ is the coefficient of the random perturbations. In the case of AGN, x corresponds to the flux
or magnitude. $W(t)$ is a Wiener process, and $dW(t)$ means a white
noise process with $\mu$ = 0 and $\sigma^2$ = 1 \citep{Kelly2014}.

\begin{equation}
  d^{1}x + \alpha_{1} x(t) = \beta_{0} dW(t)
\label{eq: drw}
\end{equation}

Its Structure Function (SF), the average difference in amplitude between points separated by a given time interval $\Delta t$, is expressed as
\begin{equation}
    {\rm SF}(\Delta t) = {\rm SF}_{\infty }(1 - e^{-\mid\Delta t\mid/\tau})^{1/2} , {\rm SF}_{\infty } = \sqrt{2}\sigma. 
\label{eq: sf}
\end{equation}
There are two key parameters: the characteristic timescale $\tau$, and the long-term variability amplitude ${\rm SF}_{\infty}$.

The second-order CARMA(2,1) or Damped Harmonic Oscillators (DHO) is also applied to simulate AGN with quasi-periodic features, such as blazars. For the detailed formula, see \citet[Table A1]{Sheng2022}.

This project used the DRW parameters derived from 7384 quasars from the Sloan Digital Sky Survey Stripe 82 field \citep{MacLeod2010}, and extended them to both DRW and DHO cases. 
The \textit{EzTao} Python Package \citep{Yu2022} was applied to simulate CARMA light curves with daily observations in the $u,g,r,i,z,y$ bands. Then, five proposed LSST cadence strategies\footnote{The five chosen cadence strategies are baseline, u\_long, filterdist, cadence\_drive and rolling. Details see \citet[section 3.2]{Sheng2022} } were selected to downsample the light curves to simulate realistic observations. On average, LSST will re-observe an object in some band every $\sim3$ days and in the same band every $\gtrsim 7$ days. These could then be modelled using an SRNN to attempt to recover the input CARMA parameters from the incomplete data.

\subsection{Representativeness of the simulated data}
It is worth noting that there are differences between CARMA models and the true AGN light curves, which may cause the former to be less representative of the latter:

\begin{enumerate}
    \item CARMA models are stationary time-series processes, but AGN light curves seem to be non-stationary \citep{Tachibana2020}.
    \item CARMA models are statistical without any physical mechanism. Moreover, quasars can have occasional large flares on top of their DRW-like variability.
    \item CARMA models do not consider the correlations between bands, whereas quasars' timescales and variability amplitude varies with bands.
\end{enumerate}

\subsection{Limitations of LSST data}

Most of the proposed WDF cadence strategies distribute unbalanced observations among the six bands, with a large allocation in the $r$, $i$, $z$, $Y$ bands and less in the $u$ and $g$ bands. Furthermore, the observations for each band are not simultaneous. 
Those factors might bring difficulties for multi-band time-series modelling, with gaps in observations leading to poor sensitivity in recovering short-timescale variability \citep{Sheng2022}.

Also problematic for the case of AGN light curve analysis with the DRW model, \citet{Kozlowski2017} suggest that reliably measuring the variability timescale $\tau$ requires a temporal baseline of at least $8-10 \tau$. With a finite duration of 10 years in the LSST survey, this introduces significant biases in recovering timescales for the large fraction of AGN with $\tau \gtrsim 1$ year, regardless of the machine learning employed \citep{Sheng2022}.

\section{Stochastic Recurrent Neural Networks high-level overview}

\begin{figure*}
\centering
\includegraphics[width=0.46\textwidth]{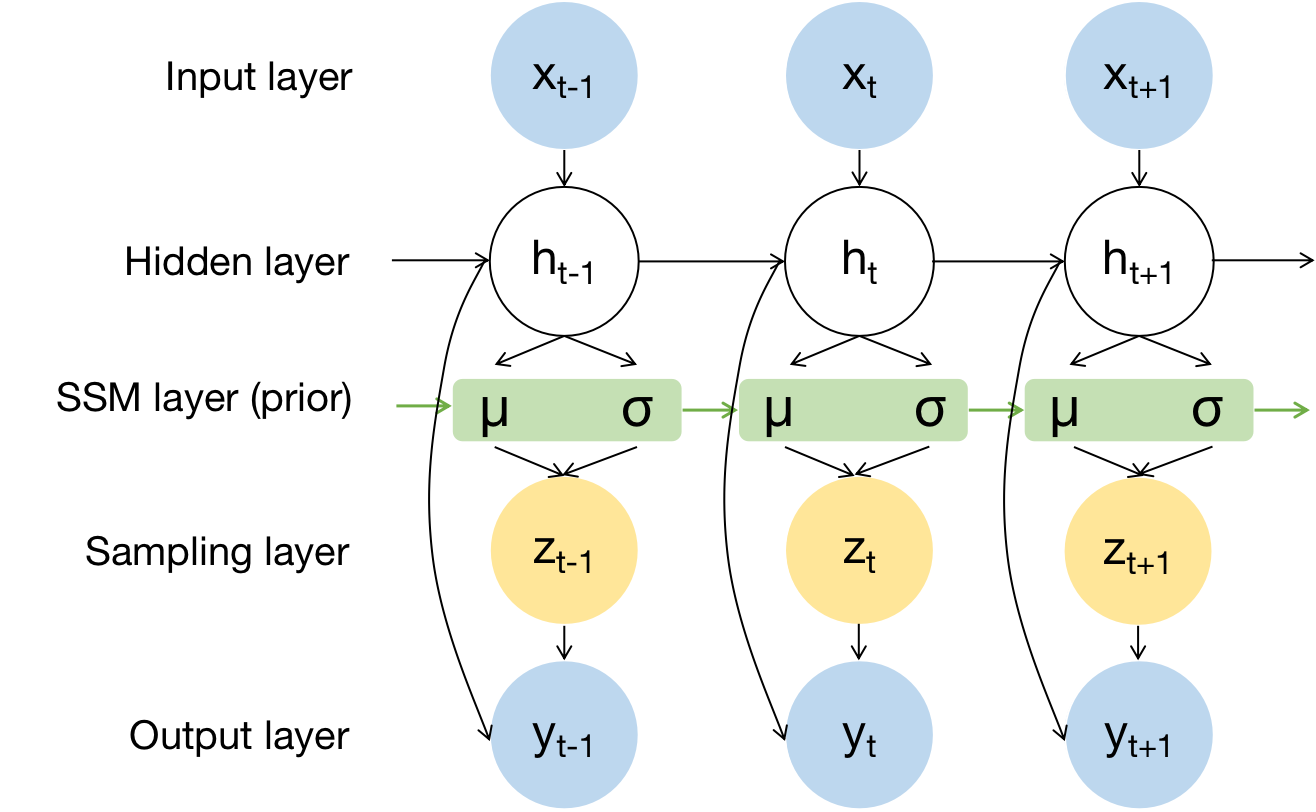}
\includegraphics[width=0.46\textwidth]{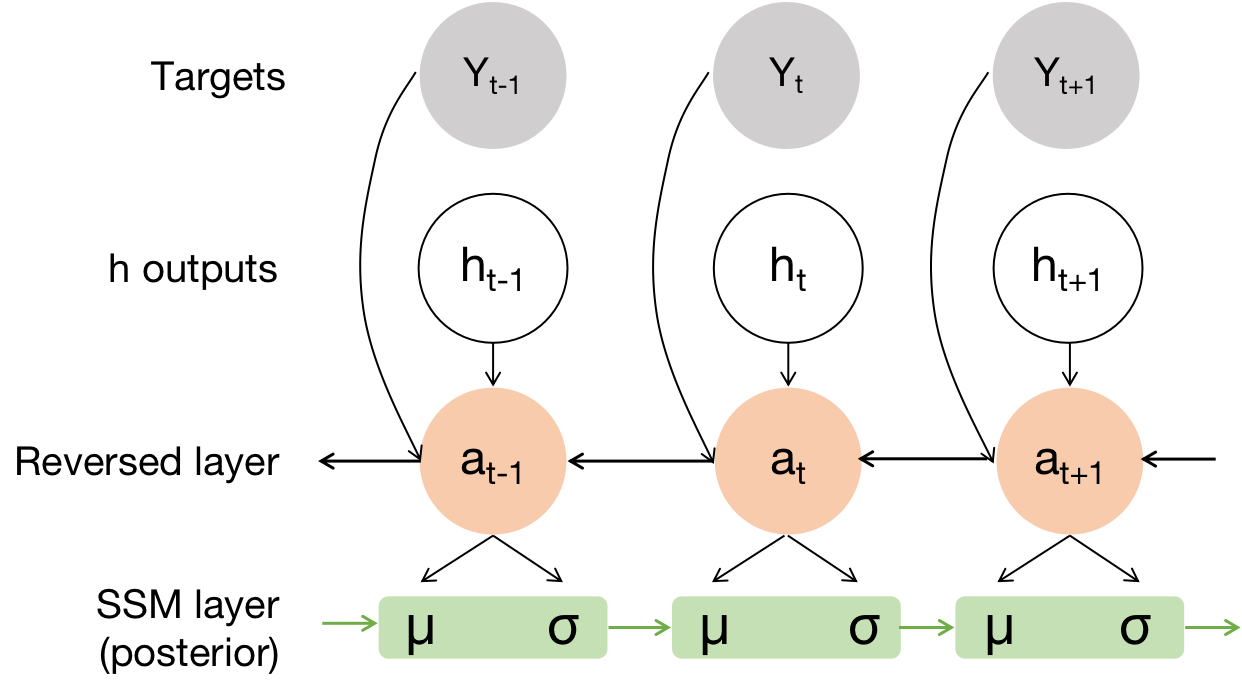}
\caption{
(a) Generative model (left) \citep[Figure 7(a)]{Sheng2022}. Observed light curves are fed into the Input layer, then a number of hidden layers (RNN layers). The output of the last hidden layer has two paths: one copy is fed to the SSM layer, realized by multiple \textbf{prior} Gaussian distributions at each time step, and then the sampling layer will randomly sample a value from each Gaussian distribution; the other copy will be combined with the sampled values and fed to the output layer. The output layer produces predicted daily light curves.
(b) Inference network (right) \citep[Figure 7(b)]{Sheng2022}. It is only used for training process. The output of the last hidden layer is combined with the target light curves at each timestep and fed into a reversed RNN layer, producing the approximate \textbf{posterior} Gaussian distributions. 
}
\label{fig:our_SRNN_model}
\end{figure*}

Inspired from \citet[][]{Justin_Christian2015},
\citet[][]{Fraccaro2016} propose the idea of adding stochasticity in a latent state representation on the classical Recurrent Neural Networks (RNN). 
They stack a state space model (SSM) on deterministic RNNs to achieve a stochastic and sequential generative model (see Figure \ref{fig:our_SRNN_model}a) and a structured variational inference network (see Figure \ref{fig:our_SRNN_model}b), which produce the output sequences and provide the model's posterior distributions, respectively. 
The loss function includes the negative log-likelihood of the predictions and targets and Kullback-Leibler divergence ($D_{KL}$) \citep[][]{Kullback1951} between the prior and posterior distributions.

\begin{figure*}
\centering
\includegraphics[width=1\textwidth]{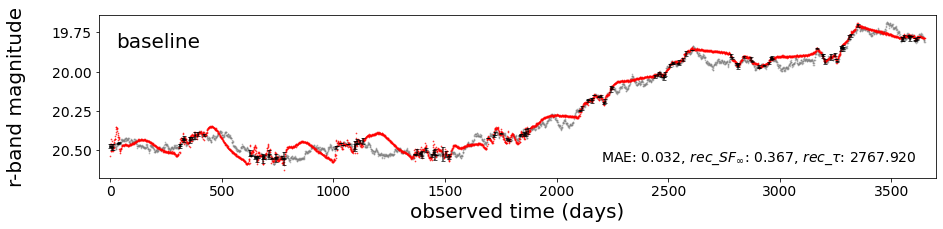}
\caption{An example of SRNN modelling. The input light curves are the observed light curves with time dilation considered, presented with grey points. Black points with error bars show the light curve with `baseline' cadence. The reconstructed light curves (by SRNN) are shown with red points. The lower left show the Mean Square Errors of the SRNN modelling and recovered DRW parameters by Gaussian Process Regression.
}
\label{fig: SRNN_lc}
\end{figure*}

This algorithm is expected to be compatible with CARMA models as CARMA can be represented as state space models. The
SRNN is applied to the simulated LSST-cadence AGN light curves, and outputs the predicted/interpolated light curves on a daily cadence over 10 years. Figure \ref{fig: SRNN_lc} is an example.

\subsection{Limitations of SRNN}
The results from \citet[Section 5]{Sheng2022} show the SRNN modelling performance for both uniformly-sampled and LSST-like light curves. 
Given the similar amounts of observation numbers, SRNN can model light curves better with uniform cadences than with LSST cadences.
SRNN can recover the long-term variability $SF_{\infty}$ well, but the timescale $\tau$ is always underestimated when $\tau$ is long, which is restricted by the number of input observations and the gaps between groups of observations.

Compared with Variational Auto-encoders, \citep[such as][]{Sanchez-Saez2021}, the SRNN also lacks explanations of latent features. The correlations between close and distant time steps are not human-interpretable. 

SRNN is designed to estimate and compare 10-year length light curves with potential cadence strategies, however, for the upcoming LSST data, SRNN modelling could be difficult as the light curves are much shorter.

\subsection{Problems of `filling the gaps'}

Here we discuss how SRNN modelling fills in the gaps between distant observations. 
Shown from \citet[Figures 8-11]{Sheng2022}, SRNN can reconstruct the input observations when cadence gaps are reasonably short compared with their timescales, but for large gaps, SRNN's general performance is weak. The following factors all affect the SRNN light curve reconstruction:

\begin{enumerate}

\item Number of observations. 

\item {Cadence strategies and different bands.}

\item {Level/timescale of variability: high ${\rm SF}_{\infty}$/short $\tau$.}

\item {Quasi-Periodicity.}

\item {Assumption of stationarity.}

\end{enumerate}

In summary, for the LSST cadences shown in \citet[Figures 9-11]{Sheng2022}, long gaps exist between observations, and for the reasons above, the SRNN model struggles to impute the behaviour during these gaps, especially for the non-periodic DRW and DHO-overdamped cases. This will turn out to be an important limitation when attempting to infer CARMA parameters from these light curves.

\section{Better data or better models?}

Recently, there has been a discussion in machine learning research: Do better data or better models contribute more to high accuracy? The answer is that for the same model, the accuracy rate increases with the amount of training data, but its `marginal utility' decreases. The capability of models are restricted by the data volume though a better model is able to improve the accuracy to a certain level.

In astronomy, the situation can be more complicated:
Data quality is not always sufficient due to weather conditions, satellite interference and other constraints. Accordingly, it is worth discussing whether we should use only "good" (quality trimmed data) or real data for training, validation, and testing.

While effective data preprocessing methods can greatly improve model results, there are some tricky tasks that developers need to be aware of. For example, how to replace missing values with values that have no significant physical meaning? How to scale, normalize and feed the input data? How to deal with poor-quality data that retains useful information? How to understand the model reflection?
Compared with designing model architectures, these issues are more prominent and deserve attention.

\section{Conclusions}
In this paper, we discuss the unavoidable problems of real and simulated astronomical data for machine learning applications as well as the limitations of applying SRNN for astronomical time series. 

However, the existing difficulties in this project are not uncommon. Researchers are expected to conduct more investigations into the model interpretation and data sets while developing machine learning algorithms and applying them to specific astronomical tasks.

\bibliographystyle{aa}
\bibliography{bonifacio}

\end{document}